\documentclass[9pt,twocolumn,twoside]{osajnl}

\journal{ol} 

\setboolean{shortarticle}{true}
\usepackage{xcolor}

\title{Josephson oscillations of edge quasi-solitons in a photonic-topological coupler}

\author[1,*]{Nataliia Bazhan}
\author[2, 3]{Boris Malomed}
\author[1]{Alexander Yakimenko}

\affil[1]{Department of Physics, Taras Shevchenko National University of Kyiv, 64/13, Volodymyrska Street, Kyiv 01601, Ukraine}
\affil[2]{Department of Physical Electronics, Faculty of Engineering, and Center
for Light-Matter Interaction, Tel Aviv University, Tel Aviv 69978, Israel}
\affil[3]{Instituto de Alta Investigaci\'{o}n, Universidad de Tarapac\'{a}, Casilla 7D, Arica, Chile}

\affil[*]{Corresponding author: nataliia.bazhan@gmail.com}

\begin{abstract}
We introduce a scheme of a photonic coupler built of two parallel topological-insulator slab waveguides with the intrinsic Kerr nonlinearity, separated by a lattice spacing. Josephson oscillations (JO) of a single edge quasi-soliton (QS) created in one slab, and of a pair of QSs created in two slabs, are considered. The single QS jumping between the slabs is subject to quick radiative decay. On the other hand, the JO of the copropagating QS pair may be essentially more robust, as one QS absorbs dispersive waves emitted by the other. The most robust JO regime is featured by the pair of QSs with phase shift $\pi$ between them.
\end{abstract}

\setboolean{displaycopyright}{true}

\begin{document}

\maketitle
Topological insulators (TIs) are materials that are insulating in a bulk but
conducting at the surface, due to the existence of scattering-resistant
topological edge states. Recently, it was discovered that the topological
phases are not restricted to solid-state fermionic states, but can also be
realized in photonic crystals and metamaterials \cite%
{TopPhot1,RevModPhys.91.015006,1912_01784,2D-3D,Szameit,microcavity}.
Rapidly growing interest in topological effects in photonics is motivated by
possibilities to design light-guiding and routing photonic circuits in a
manner that is stable against disorder, due to the robustness of topological
edge states. Developing these studies, solitons have been recently observed
in the bulk of a photonic Floquet TI \cite{mukherjee2020observation}, and
theoretically elaborated in other TI setups \cite{Konotop1,Konotop2,Konotop3}%
. They exhibit dynamics different from that demonstrated by solitons in
ordinary bandgap settings \cite{KA}, \textit{viz}., cyclotron-like orbits
induced by the photonic-lattice topology. The concept of topological lasers
based on the Floquet TI was put forward in \cite{Photonics_4_126101_2019}.
It is based on a truncated array of lasing helical waveguides, with the
pseudo-magnetic field induced by their twist along the propagation
direction, opening up a topological lattice gap by breaking the
time-reversal symmetry. Localized edge states in a similar system were
analyzed too \cite{PhysRevA.90.023813}. Further, the recent study \cite%
{doi:10.1021/acsphotonics.0c01771} dealt with topological edge states
maintained by a domain wall between two helical honeycomb lattices with
opposite helicities. In that system, nonlinearity helps to create robust
edge states in the form of fundamental and multipole solitons, including
moving ones.

In this work, we address interactions between nonlinear topological surface
wave packets copropagating along opposite edges of two slab-shaped TI
waveguides separated by free space. The setup, displayed in Fig. \ref%
{fig:3d-lattice}, is a \textit{coupler}, i.e., a set of two parallel
waveguides with an empty space between them, which are \textit{coupled} by
tunneling of the field. Oscillatory dynamics of wave modes in the coupler is
usually categorized as \textit{Josephson oscillations} (JO) \cite%
{coupler-book}. Long Josephson junctions \cite{Ustinov} were the first
well-studied example of nonlinear couplers for wave modes (fluxons in bulk
superconductors separated by a narrow dielectric barrier). Various
realizations of couplers are also known in optics \cite{coupler-optics} and
Bose-Einstein condensates \cite{coupler-BEC,Smerzi,Markus, Viskol}. The
present work aims to elaborate the scheme of the \textit{photonic TI coupler}
and analyze the dynamics of composite nonlinear modes in it. In this
connection, it is relevant to mention a recently elaborated scheme for
resonant coupling between excitations on opposite edges of a single TI slab
\cite{Kartashov}.

The photonic structure of each waveguiding slab building the coupler is
similar to that proposed in Refs. \cite%
{PhysRevLett.117.143901,PhysRevLett.117.013902} for the creation of surface
solitons in the semi-infinite bulk waveguide. The band structure of a
single-slab lattice is similar to the band structure of anomalous Floquet
TIs investigated in \cite{PhysRevLett.117.143901}. The 2D Floquet TI
maintains protected edge states even if all bands have zero Chern number
\cite{PhysRevB.82.235114, PhysRevX.3.031005, PhysRevLett.114.056801,
PhysRevLett.114.106806}. Our consideration reveals the propagation of
nonlinear wave packets at the edge of the lattice, in agreement with the
topological structure of the underlying linear band structure described in
\cite{PhysRevLett.117.013902}. In the paraxial approximation, the
propagation of the optical-beam envelope $\psi (x,y,z)$ along coordinate $z$
obeys the nonlinear Schr\"{o}dinger equation (NLSE),
\begin{equation}
i\partial _{z}\psi =-\left( 2k_{0}\right) ^{-1}\nabla _{\perp }^{2}\psi
-k_{0}n_{0}^{-1}\Big(n_{L}(x,y,z)+n_{2}|\psi |^{2}\Big)\psi ,
\label{eq:nlse}
\end{equation}%
with the diffraction operator, $\nabla _{\perp }^{2}\equiv \partial
_{x}^{2}+\partial _{y}^{2}$, acting on transverse coordinates $\left(
x,y\right) $. Here $k_{0}=2\pi n_{0}/\lambda $ is the wavenumber
corresponding to the carrier wavelength $\lambda $, $n_{0}$ is the
background refractive index, $n_{2}|\psi |^{2}$ is the nonlinear correction
to it, and $n_{L}(x,y,z)$ represents the helix-lattice background \cite%
{oe-27-1-121} with helix radius $R_{0}$, lattice spacing $a$, and modulation
period $Z$. We adopt parameters consistent with fused silica glass at $%
\lambda =1.55\,\mu $m: $n_{0}=1.45$, $n_{L}=2.7\times 10^{-3}$ in the
waveguides (and $n_{L}=0$ outside), and $n_{2}=3\times 10^{-7}$ cm$^{2}$/GW
\cite{2006OExpr..14.6055S}. Individual helices have the circular cross
section with $R_{0}=4\,\mu $m. The beam intensity $|\psi |^{2}$ is
normalized by a characteristic value $I_{0}=10^{3}$ GW/cm$^{2}$, for which
the nonlinear index shift $n_{2}I_{0}$ is comparable to $n_{L}$. For the
modal cross-section area of the helix $w_{0}^{2}=(10$ $\mu $m)$^{2}$, this $%
I_{0}$ corresponds to peak powers $\sim 1$ MW, readily accessible with
pulsed lasers~\cite{2006OExpr..14.6055S}. Here we do not consider temporal
effects, which is a subject for a separate work.

\begin{figure}[tbh]
\caption{A sketch of two photonic lattices composed of helical waveguides.
The lattices are separated by the spacing of width $\Delta $. One lattice
consists of (green) waveguides twisted clockwise, while the other (orange)
lattice is the mirror reflection of the green one relative to the midplane
of the lattice spacing. This setup allows copropagation of two QSs
(quasi-solitons) on opposite edges of the lattice spacing. The waveguides in
each lattice are shifted relative to each other along the $z$-axis by half
of the modulation period, $Z=1$ cm. Shown are the lattices composed of $%
n_{y}^{s}=4$ shells ($n_{y}=2$).}\centering
\includegraphics[width=0.9\linewidth]{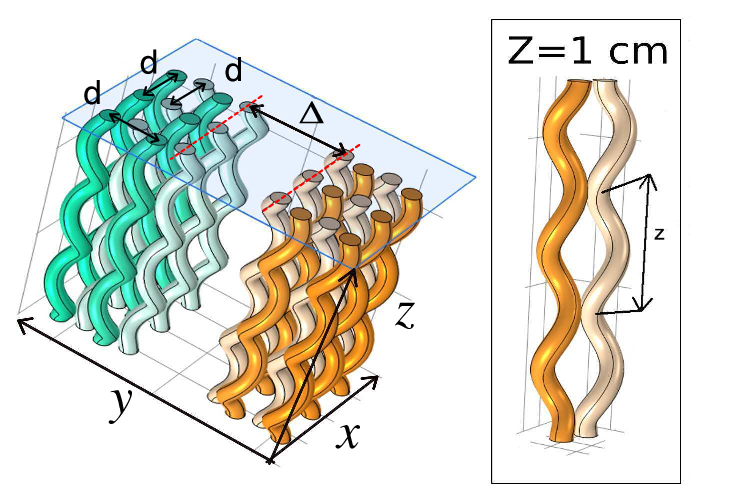}
\label{fig:3d-lattice}
\end{figure}

In this work we consider a coupler composed of two mirror-symmetric
staggered helical lattices separated by a gap of width $\Delta $, as shown
in Fig. \ref{fig:3d-lattice}. Due to opposite signs of the twist of the
waveguides in the lattices, two edge nonlinear wave packets copropagate in
the same direction along the edges of the lattices, and interact with each
other across the spacing. For each lattice we use parameters of nonlinear
photonic Floquet TIs introduced in \cite%
{PhysRevLett.117.143901,PhysRevLett.117.013902}. The lattices consist of 2D
square arrays of helical waveguides, \textit{staggered} so that adjacent
waveguides have helix phase shifts of $\pi $ relative to each other, as
shown in the right panel of Fig. \ref{fig:3d-lattice}. Note that, while $z$
varies within one period $Z$, each waveguide is approaching its four
neighbors sequentially. The respective modulation of the linear refractive
index in Eq. (\ref{eq:nlse}), which describes the coupler shown in Fig. \ref%
{fig:3d-lattice}, is adopted as
\begin{align}
n_{L}(x,y,z)& =\Delta n_{1}\sum_{n=-[n_{x}/2]}^{[n_{x}/2]}\sum_{m=1}^{n_{y}}%
\Big\{V_{0}(x-X_{n}^{-},y\mp Y_{m-1}^{\pm })+  \notag \\
& +V_{0}(x-X_{n-1/2}^{+},y\mp Y_{m-1/2}^{\pm })\Big\},  \label{eq:lp}
\end{align}%
where $n_{x}\,$ and $n_{y}$ are numbers of shells, equal in both lattices
(accordingly, total numbers of waveguiding rows and columns in each slab are
$n_{y,x}^{s}=2n_{y,x}$). We build the coupler starting from its center.
Thus, we summation in Eq. (\ref{eq:lp}) includes all waveguides along the $x$
axis in the range of $\left\{ -[n_{x}/2],[+[n_{x}/2]\right\} $, where $[...]$
stands for the integer part. Both slabs include $n_{y}$ shells, factors $%
V_{0}(y-Y_{m}^{+})$ and $V_{0}(y+Y_{m}^{-})$ pertaining to the top and
bottom slabs. Further, functions introduced in Eq. (\ref{eq:lp}) are
\begin{equation}
X_{n}^{\pm }(z)=nd\pm x_{0}(z),\,\,\,\,Y_{m}^{\pm }(z)=\Delta /2+md\pm
y_{0}(z).  \label{X}
\end{equation}%
Here, $\Delta n_{1}=2.7\times 10^{-3}$ is the modulation depth, $d=\sqrt{2}a$
is the lattice constant and $\Delta $ the spacing width in Fig. \ref%
{fig:3d-lattice}.

We assume a hypergaussian waveguide's profiles in Eq. (\ref{eq:lp}), \textit{%
viz}.,
\begin{equation}
V_{0}(x,y)=\exp \left( -[(x^{2}+y^{2})/\sigma ^{2}]^{3}\right) ,  \label{V0}
\end{equation}%
with $\sigma =4\,\mu $m. The waveguides' axes are shaped as helices,
\begin{equation}
x_{0}(z)=R_{0}\cos (\Omega z),\quad y_{0}(z)=R_{0}\sin (\Omega z),
\label{x0y0}
\end{equation}%
with radius $R_{0}=4\,\mu $m and pitch $Z=2\pi /\Omega =1$\thinspace cm.
These parameters are chosen to provide a balance between the minimization of
bending losses, which inevitably accompany the nearest-neighbor couplings in
the present setting, and the necessity to fit sufficiently many helix cycles
to an experimentally feasible total array length, $\lesssim 10$~cm.

\begin{figure}[tbh]
\centering
\includegraphics[width=\linewidth]{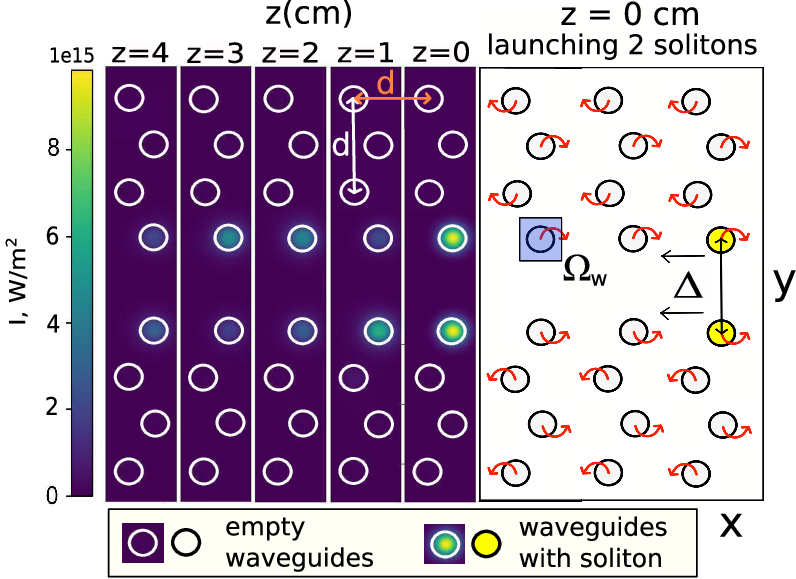}
\caption{Configurations of the copropagating QSs in the dual-waveguide
coupler in the $\left( x,y\right) $ cross-section at different values of
propagation distance $z$. The QSs are launched at $z=0$, and subsequent
configurations, produced by the numerical solution of Eq. (\protect\ref%
{eq:nlse}), are displayed at $z=1,2,3,4\,$cm. The distance between
waveguides in each lattice is $d\approx 31$ $\protect\mu $m. The spacing
between the lattices, $\Delta $, is the same as in Eq. (\protect\ref{X}).
Circles indicate cross-sections of individual waveguides and arrows
designate the direction of their twist. See Visualization 1 in the
Supplement as an example.}
\label{fig:xy-lattice}
\end{figure}

Our objective is to explore the interaction between two wave packets
copropagating along edges of the slab-shaped lattice waveguides forming the
coupler in Fig. \ref{fig:3d-lattice}. We consider the nonlinear edge packets
as ``quasi-solitons" (QSs), as they keep soliton-like shapes, while
radiating power at a small rate \cite{mukherjee2020observation}. As shown by
Fig. \ref{fig:xy-lattice}, the QSs move in the same direction due to the
mutual mirror symmetry of the parallel slabs. The interaction between them
is mediated by electromagnetic field traversing the spacing between the
slabs.

\begin{figure}[tbh]
\centering
\includegraphics[width=\linewidth]{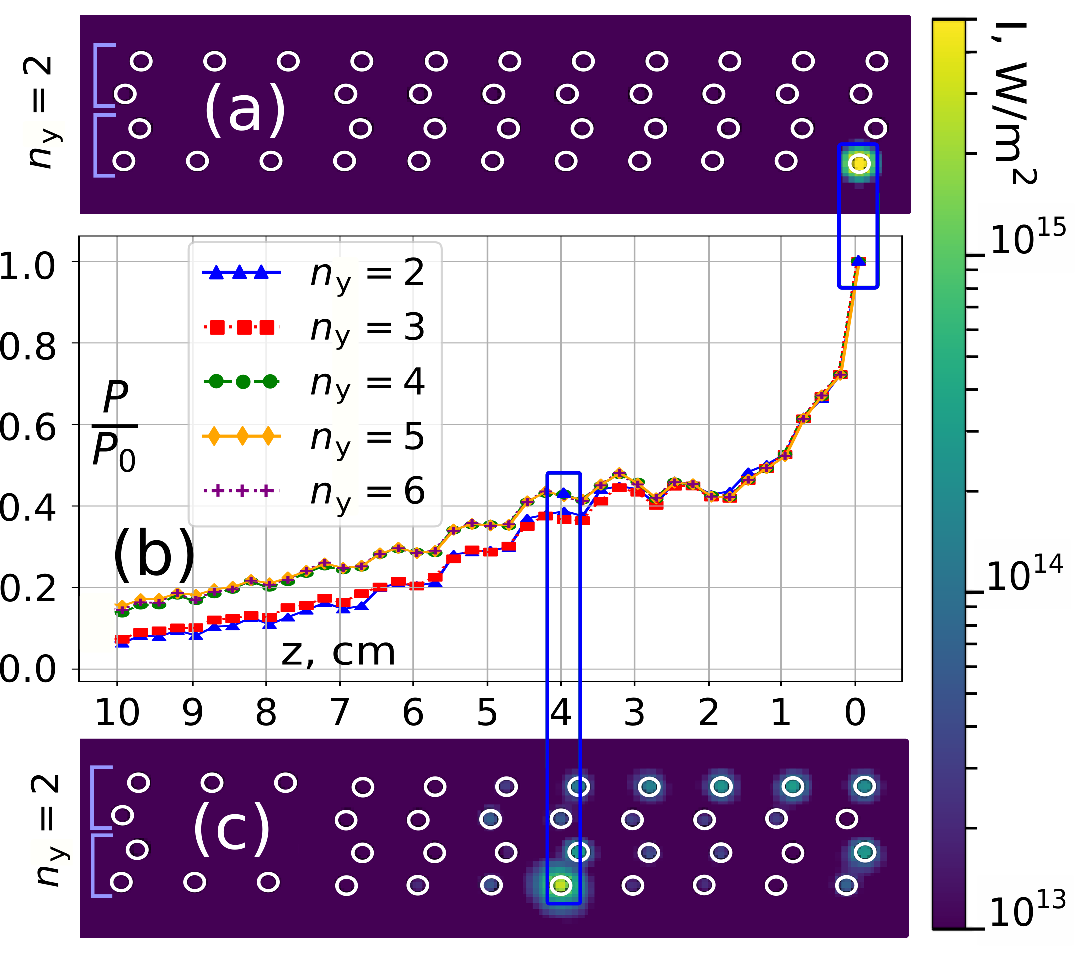}
\caption{The single edge QS traveling from right to left in the lattice slab
built of $n_{y}$ layers. In (a) and (c) yellow circles show the location of
the QS at $z=0$ and $z=4$\thinspace cm, respectively. The logarithmic color
bar on the right indicates the intensity. White circles indicate
cross-sections of individual helix waveguides. In (c), the energy lost by
the quasi-soliton spreads along the lattice waveguides. (b) The evolution of
the peak power of the QS.}
\label{fig:intensity_evolution}
\end{figure}

First, we produce individual edge QS solutions of NLSE (\ref{eq:nlse}) with
the effective potential defined as per Eqs. (\ref{eq:lp})-(\ref{x0y0}),
starting from input
\begin{equation}
\Psi \left( x,y;z=0\right) =\sqrt{I}\exp (-(\mathbf{r}-\mathbf{r}%
_{0})^{2}/2a^{2}),  \label{initial}
\end{equation}%
\noindent where $\mathbf{r}_{0}$ is the initial position of the QS, and $I$
is its peak power. The edge QS, while it can pass a considerable distance,
is not a completely stable object, being subject, as mentioned above, to
losses due to emission of small-amplitude dispersive waves \cite%
{PhysRevLett.117.143901}, see Fig. \ref{fig:intensity_evolution}. The loss
rate depends on number $n_{y}$ of layers in the lattice slab in the
direction across the lattice, see Fig. \ref{fig:intensity_evolution}. To find an
optimal value of $n_{y}$, we computed the power of a single QS propagating
along the edge of one slab. The results, displayed in the top plot of Fig. %
\ref{fig:intensity_evolution}, demonstrate that the edge QS traveling in a
thin slab, with $n_{y}\leq 3$, decays faster than in the thicker one, with $%
n_{y}\geq 4$. Therefore, we ran the systematic numerical analysis using the
slabs with $n_{y}=4$. Then, Fig. \ref{fig:intensity_evolution} shows that
the peak power of the QS decreases by $\simeq 10$ times, having passed $10$
cm in the $z$ direction. Because the QS moves along the $z$ axis in the
lattice with pitch $Z=1$ cm, it shifts by $10$ helix waveguides in the $x$
direction while passing $10$ cm along $z$. This means that one needs to have
the number of layers $n_{x}\geq 10$ in the slab along $x$. In the
simulations, we used the lattices with $n_{x}=18$. Animations of QSs
propagating on the opposite edges of the coupler are presented in
Supplemental Material.

To address the interaction of two edge QSs carried by the parallel slabs in
the coupler, as shown in Fig \ref{fig:xy-lattice}, we assume that they move
with equal velocities, which maximizes the impact of the interaction.
Generally, the interaction effects are transient ones because, as mentioned
above, each QS in isolation loses $90\%$ of its energy after passing the
distance $z=10$ cm. The dominant mechanism of the losses is the diffraction
of the wave packet, while bending losses are very low \cite%
{PhysRevLett.117.143901,PhysRevLett.117.013902}. It is well known that the
interaction between solitons depends on their peak powers, distance, and
relative phase, $\Delta \phi $ \cite{RMP}.
We monitored the evolution of $\Delta \phi (z)=\phi _{top}-\phi _{bott}$ and the power difference between the edge QSs maintained by
the top and bottom slabs, $\Delta P(z)=\int \int_{\Sigma }\left[ \left\vert
\psi _{top}(x,y,z)\right\vert ^{2}-\left\vert \psi _{bott%
}(x,y,z)\right\vert ^{2}\right] dxdy$, where $\Sigma $ is the biggest area
around the given helical waveguide which does not touch adjacent ones, see
the right panel of Fig. \ref{fig:xy-lattice}), while $\phi _{top}$
and $\phi _{bott}$ are the phases measured at centers of the
waveguides.

The power distribution in the interacting QSs depends on spacing $\Delta $,
the initial value of $\Delta \phi $, and the initial power (taken equal for
both solitons). As the result of the interaction, an essential part of the
QSs' energy oscillates between the slabs, realizing the JO with the
frequency which depends on $\Delta $ and $\Delta \phi $. Josephson
oscillations are observable only between close enough slabs.

\begin{figure}[tbh]
\centering
\includegraphics[width=\linewidth]{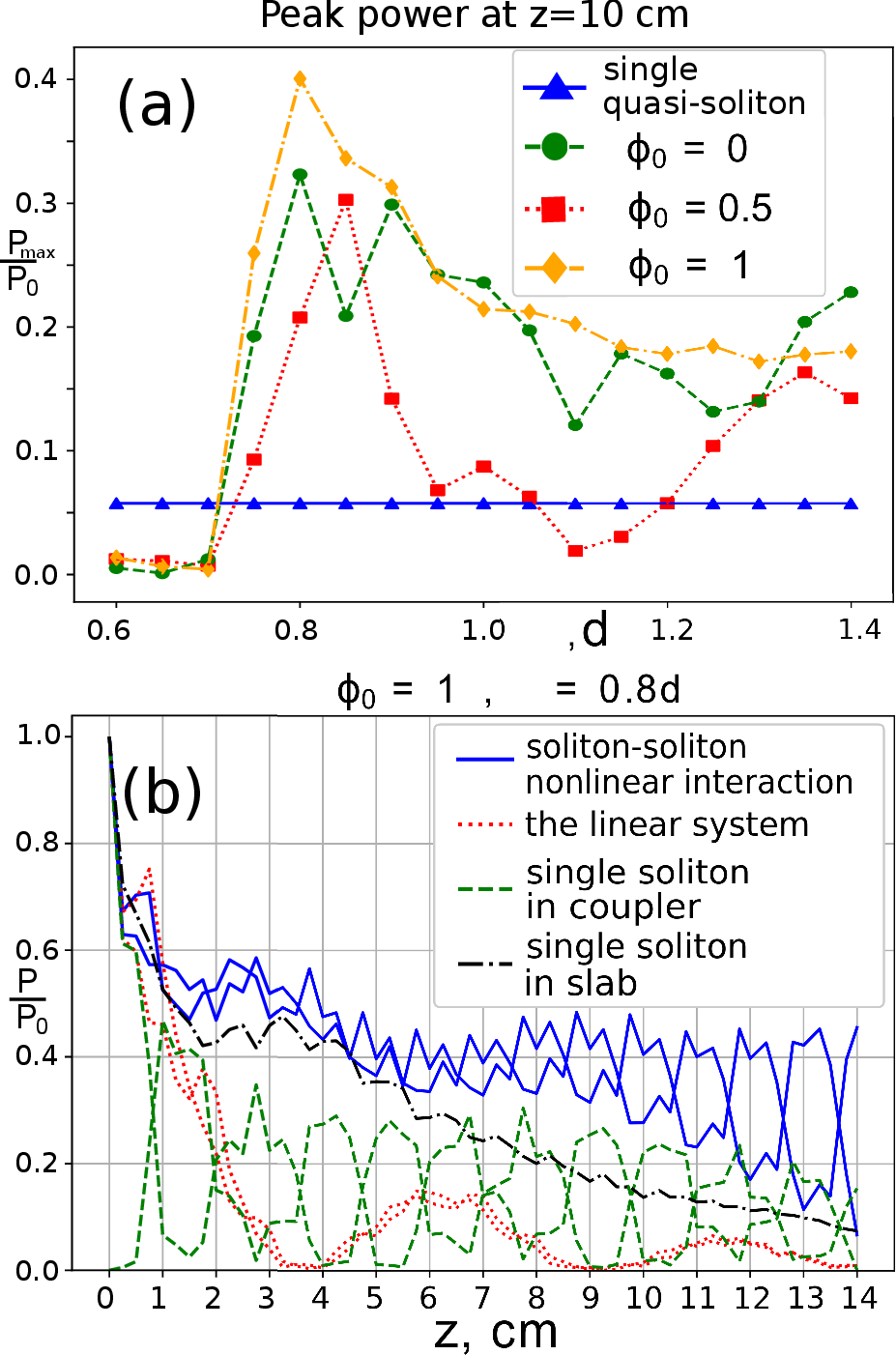}
\caption{(a) The peak power (normalized to $P_{0}$, see Eq. (\protect\ref%
{initial})) of the single QS and QS pair, with different phase shifts, after
having passed distance $z=10$ cm, vs. the spacing between the parallel
waveguiding slabs. (b) The peak power of the single QS, and largest powers
of the QS pair in the top and bottom slabs, vs. the propagation distance,
for the coupler with spacing $\Delta =0.8d$ and phase shift of the pair $%
\Delta \protect\phi (0)=\protect\pi $ (it is the optimal set of parameters
for the realization of the JO by the pair, according to panel (a)). Also
displayed is the decay of the same initial pair of pulses as the QS, but in
the linear system. See Visualizations 1-4 in the Supplement illustrating the
graph curves.}
\label{fig:increasing-lifedistance}
\end{figure}

A significant result produced by systematic simulations of the JO is that
the two-QS configuration is essentially \emph{less vulnerable} to the decay
than the single edge QS. This trend is explained by the fact that dispersive
waves emitted by one QS are absorbed by the other. Accordingly, in the
two-QS dynamical regimes the energy stays concentrated in the QSs, making
them brighter.

In Fig. \ref{fig:increasing-lifedistance}(a), the peak power of the single-
and two-QS configurations is displayed as a function of the spacing, $\Delta
$, in the state produced by the simulations at $z=10$ cm. It is seen that
the strong interaction across the narrow spacing, with $\Delta \leq 0.75d$,
leads to fast decay of the QSs, faster than the single QS loses its energy.
However, at $\Delta =0.80d$, the interaction gives a significant boost to
the QS's robustness for all initial phase shifts, the largest boost being
observed for $\Delta \phi (0)=\pi $. The latter finding is natural, as the
repulsion between the QSs helps to suppress their decay.

In Fig. \ref{fig:increasing-lifedistance}(b) we compare the results produced
by the propagation of the single edge QS on a separate slab, JO of a single
edge QS in the coupler, and the pair of QSs with $\Delta \phi (0)=\pi $ in
the coupler with $\Delta =0.80d$ (the latter configuration provides the most
robust QSs). The results, represented by the dependence of the peak power on
the propagation distance, $z$, clearly demonstrate that the interaction
between the QSs indeed helps to stabilize them, due to the mutual absorption
of the dispersive waves emitted by each one. For the comparison's sake, Fig. %
\ref{fig:increasing-lifedistance}(b) also displays very fast decay of the
same pair of pulses as the initial QSs, but in the linear system. In
Supplemental Material we provide videos of the system's evolution for cases
illustrated in Fig. \ref{fig:increasing-lifedistance}(b), as well as other
details of the JO dynamics.

The topology of the coupler plays a key role in protecting the edge QSs from
decay when passing the defects. It can be demonstrated (see details in
Supplement) that missing edge waveguides or even the absence of extended
segments of the edge lattice does not destroy the coupled evolution of the
QSs pair (see Visualizations 5, 7-10 in the Supplement). On the other hand,
a missing waveguide near the edge of the single slab causes decay of a QS
passing the defect (see Visualization 2). The only situation in which the
QSs remain robust in the presence of the defect is when they jumps over it
due to JO in the two-slab coupler (see Visualizations 7-13).



In conclusion, we have analyzed the dynamics of the single edge QS
(quasi-soliton) and a pair of QSs copropagating in the coupler built of
parallel photonic TIs (topological insulators) separated by space. The
single QS performs JO\ (Josephson oscillations) between the parallel
waveguides, quickly losing energy through the emission of radiation. The
evolution of the QS pair demonstrates essentially more robust dynamics, as
dispersive waves emitted by one QS are absorbed by the other, in the course
of their coupled JO. The two-QS dynamics depends on the phase shift $\Delta
\phi (0)$ between the QSs, the most robust regime corresponding to $\Delta
\phi (0)=\pi $, in which case they periodically bounce back from each other.
As an extension of the analysis, it may be interesting to consider a
circular cavity, in the form of the photonic TI (topological-insulator)
coupler closed into a ring. The JO of the QSs running along the ring may be
additionally stabilized by external gain.

\textbf{Acknowledgments}.The authors thank O. G. Chelpanova and V. R. Tuz
for fruiful discussions.

\textbf{Funding}.The work of B.A.M. was supported, in part, by the Israel
Science Foundation, through grant No. 1286/17. NB and AY acknowledge support
from National Research Foundation of Ukraine, through grant No. 2020.02/0032.

\textbf{Disclosures}. The authors declare no conflicts of interest.
\textbf{Data availability}. No data were generated or analyzed in the presented research.

See Supplement 1 for supporting content.

\bibliography{Topological_Refs}
\cleardoublepage
\begin{enumerate}
  \item Soljaci Marin Lu Ling Joannopoulos John D. “Topological photonics”. Nature Photonics8.11 (Nov. 2014), pp. 821–829.
  \item Tomoki Ozawa et al. “Topological photonics”. Rev. Mod. Phys.91(1 Mar. 2019), p. 015006.
  \item Daria Smirnova et al. “Nonlinear topological photonics”. Appl. Phys.7 (2020), p. 021306.
  \item Rho Junsuk Kim Minkyung Jacob Zubin. “Recent advances in 2D,3D and higher-order topological photonics”. Light: Science andApplications9 (July 2020), p. 130.
  \item Mark Kremer et al. “Topological effects in integrated photonic waveguide structures”. Opt. Mater. Express11.4 (Apr. 2021), pp. 1014–1036.
  \item Dmitry D. Solnyshkov et al. “Microcavity polaritons for topological photonics”. Opt. Mater. Express11.4 (Apr. 2021), pp. 1119–1142.
  \item Sebabrata Mukherjee and Mikael C. Rechtsman. “Observation of Topological Band Gap Solitons”. Science368 (2020), pp. 856–859.
  \item Sergey K. Ivanov and Yaroslav V. Kartashov and Lukas J. Maczewsky and Alexander Szameit and Vladimir V. Konotop. "Edge solitons in Lieb topological Floquet insulator". Opt. Lett (Mar. 2020).
  \item S. K. Ivanov and Y. V. Kartashov and L. J. Maczewsky and A. Szameit and V. V. Konotop. "Bragg solitons in topological Floquet insulators". Opt. Lett (Apr. 2020).
  \item Ivanov, Sergey K. and Kartashov, Yaroslav V. and Heinrich, Matthias and Szameit, Alexander and Torner, Lluis and Konotop, Vladimir V. "Topological dipole Floquet solitons". Phys. Rev. A, 2021.
  \item Y S. Kivshar and G P. Agrawal. “Optical Solitons: From Fibers to Photonic Crystals”. Academic Press(2003).
  \item Sergey K. Ivanov et al. “Floquet topological insulator laser”. APLPhotonics4 (2019), p. 126101.
  \item Ablowitz, Mark J. and Curtis, Christopher W. and Ma, Yi-Ping. "Linear and nonlinear traveling edge waves in optical honeycomb lattices". Phys. Rev. A (Aug. 2014), p. 023813.
  \item Zhiwei Shi et al. “Topological Edge States and Solitons on a Dynamically Tunable Domain Wall of Two Opposing Helical Waveguide Arrays”. ACS Photonics8.4 (2021), pp. 1077–1084.
  \item B.  A.  Malomed. "Spontaneous  Symmetry  Breaking,  Self-Trapping, and Josephson Oscillations". Springer-Verlag: Berlin and Heildelberg, 2013.
  \item A. V. Ustinov. "Solitons in Josephson Junctions: Physics of Magnetic Fluxons in Superconducting Junctions and Arrays". Pearson, NewYork, 2015.
  \item B. A. Malomed. "A variety of dynamical settings in dual-core nonlinearfibers", Handbook of Optical Fibers. Vol. 1. G.-D. Peng, Editor : Springer, Singapore, 2019, pp. 421–474.
  \item Immanuel Bloch, Theodor W. Hänsch, and Tilman Esslinger. “Atom Laser with a cw Output Coupler”. Phys. Rev. Lett.82 (Apr. 1999),pp. 3008–3011.
  \item S. Raghavan et al. “Coherent oscillations between two weakly coupled Bose-Einstein condensates: Josephson effects, $\pi$-oscillations, and macroscopic quantum self-trapping”. Phys. Rev. A59 (Jan. 1999),pp. 620–633. 
  \item Michael Albiez et al. “Direct Observation of Tunneling and NonlinearSelf-Trapping in a Single Bosonic Josephson Junction”. Phys. Rev.Lett.95 (June 2005), p. 010402.
  \item Zhaopin Chen, Yongyao Li, and Boris A. Malomed. “Josephson oscillations of chirality and identity in two-dimensional solitons in spin-orbit coupled condensates”. Phys. Rev. Research2 (Aug. 2020),p. 033214.
  \item Zhang, Yiqi and Kartashov, Yaroslav V. and Zhang, Yanpeng and Torner, Lluis and Skryabin, Dmitry V. "Resonant Edge-State Switching in Polariton Topological Insulators".  Laser \& Photonics Reviews, 2018.
  \item Daniel Leykam and Y. D. Chong. “Edge Solitons in Nonlinear Photonic Topological  Insulators”.  Phys.  Rev.  Lett.117  (14  Sept.  2016),p. 143901. 
  \item Daniel Leykam, M. C. Rechtsman, and Y. D. Chong. “Anomalous Topological Phases and Unpaired Dirac Cones in Photonic Floquet Topological Insulators”. Phys. Rev. Lett.117 (1 June 2016), p. 013902.
  \item Kitagawa, Takuya and Berg, Erez and Rudner, Mark and Demler, Eugene. "Topological characterization of periodically driven quantum systems". Phys. Rev. B (Dec. 2010), p. 235114.
  \item Rudner, Mark S. and Lindner, Netanel H. and Berg, Erez and Levin, Michael. "Anomalous Edge States and the Bulk-Edge Correspondence for Periodically Driven Two-Dimensional Systems". Phys. Rev. X (Jul, 2013), p.031005.
  \item Titum, Paraj and Lindner, Netanel H. and Rechtsman, Mikael C. and Refael, Gil. "Disorder-Induced Floquet Topological Insulators". Phys. Rev. Lett. (Feb, 2015), p. 056801
  \item Carpentier, David and Delplace, Pierre and Fruchart, Michel and Gawedzki, Krzysztof. "Topological Index for Periodically Driven Time-Reversal Invariant 2D Systems". Phys. Rev. Lett. (Mar, 2015), p.106806.
  \item Zhiwei Shi et al. “Generation and probing of 3D helical lattices with tunable helix pitch and interface”. Optics Express27.1 (Jan. 2019),p. 121.
  \item Alexander Szameit et al. “Two-dimensional soliton in cubic fs laser written waveguide arrays in fused silica”. In:Optics Express14.13(June 2006), pp. 6055–6062.
  \item Yuri S. Kivshar and Boris A. Malomed. “Dynamics of solitons in nearly integrable systems”. In:Rev. Mod. Phys.61 (Oct. 1989), pp. 763–915.
\end{enumerate}
\end{document}